\documentclass{article}

\usepackage{arxiv}

\usepackage[utf8]{inputenc} 
\usepackage[T1]{fontenc}    
\usepackage{hyperref}       
\usepackage{url}            
\usepackage{booktabs}       
\usepackage{amsfonts, amsmath}       
\usepackage{nicefrac}       
\usepackage{microtype}      
\usepackage{lipsum}
\usepackage{graphicx}
\graphicspath{ {./images/} }

\title{AI agents can coordinate beyond human scale}

\author{
    Giordano De Marzo \\
    University of Konstanz, Germany\\
    Centro Ricerche Enrico Fermi, Italy\\
    Complexity Science Hub Vienna, Austria\\
   \And
 Claudio Castellano \\
  Istituto dei Sistemi Complessi (ISC-CNR), Italy\\
  Centro Ricerche Enrico Fermi, Italy\\
  \And
 David Garcia \\
    University of Konstanz, Germany\\
    Complexity Science Hub Vienna, Austria\\
  \texttt{david.garcia@uni-konstanz.de} \\
}

\begin{document}
\maketitle
\begin{abstract}
Large language models (LLMs) are increasingly deployed in collaborative tasks involving multiple agents, forming an ''AI agent society" where agents interact and influence one another. Whether such groups can spontaneously coordinate on arbitrary decisions without external influence—a hallmark of self-organized regulation in human societies—remains an open question. Here we investigate the stability of groups formed by AI agents by applying methods from complexity science and principles from behavioral sciences. We find that LLMs can spontaneously form cohesive groups, and that their opinion dynamics is governed by a majority force coefficient, which determines whether coordination is achievable. This majority force diminishes as group size increases, leading to a critical group size beyond which coordination becomes practically unattainable and stability is lost. Notably, this critical group size grows exponentially with the language capabilities of the models, and for the most advanced LLMs, it exceeds the typical size of informal human groups. Our findings highlight intrinsic limitations in the self-organization of AI agent societies and have implications for the design of collaborative AI systems where coordination is desired or could represent a treat.
\end{abstract}

Large Language Models (LLMs) have proven individual capabilities for a wide range of applications, such as summarization \cite{chang2023booookscore}, sentiment analysis \cite{miah2024multimodal, aroyehun2023leia}, scientific research \cite{boiko2023autonomous} or mathematical reasoning \cite{romera2024mathematical}. 
Agents driven by LLMs can be used in group settings where several agents interact with each other in collaborative tasks \cite{guo2024embodied, liu2023dynamic, shen2024hugginggpt}.
Collaboration setups where multiple LLMs have different roles and tasks, such as AutoGPT \footnote{https://github.com/Significant-Gravitas/AutoGPT}, Microsoft's AutoGen \cite{wu2023autogen} or OpenAI SWARM\footnote{https://github.com/openai/swarm}, are qualitatively different from ensembling techniques, where interaction between different models is absent~\cite{jiang2023llm}. The interest for Agentic AI is on the rise, as also testified by the recent release of the Agent2Agent (A2A) protocol by Google \footnote{https://github.com/google/A2A}, with the aim of simplify the integration of several AI agents from different frameworks.
Recent advancement in on-device LLMs is also leading to AI-powered devices and assistants, such as Siri\footnote{https://openai.com/index/openai-and-apple-announce-partnership/} or the Humane AI Pin, that can perform everyday tasks in interaction with each other, for example coordinating events or negotiating prices.
\par As we move towards a society where AI agents interact with each other on our behalf, it becomes important to understand their ability to coordinate decisions in large groups.
This can motivate new applications, but also help identify risks stemming from undesired collective behavior. 
For example, trading bots interacting through the stock market can lead to flash crashes \cite{johnson2013abrupt} that motivate regulation like trading curbs.  
Current research on the behavior of LLMs has mostly focused on their behavior in isolation \cite{aher2023using, argyle2023out, dentella2023systematic, binz2023using, pellert2023ai, strachan2024testing} and collective behavior has been explored less \cite{grossmann2023ai, bail2024can, lu2024generative}, mostly with a focus on social simulation of network structures \cite{de2023emergence, papachristou2024network, chang2024llms}, opinion and information spreading \cite{park2023generative, chuang2023simulating, cau2025language} and online interaction \cite{tornberg2023simulating, park2022social, rossetti2024social}. 
To assess the reliability and trustworthiness of large numbers of interacting AI agents, we need to understand if they can form stable groups, display emerging coordination, what determines the abilities of AI agents to coordinate, and at what scale this can happen.

Group coherence is related to coordination, as the agreement on shared decisions or norms creates a social contract that regulates behavior \cite{dunbar1999culture}.
Taking quick collective decisions is often a necessity, even in situations where there is no information about the quality or utility of any option. This is the case, for example, of animal coordination when moving collectively and escaping predators \cite{couzin2005effective}.
Across species, this leads to a scaling of the average group size with brain structure, with human groups reaching sizes between 150 and 300 \cite{dunbar1992neocortex, dunbar1998social}, as documented by archaeological records \cite{casari2018group} and contemporary experience \cite{zhou2005discrete}.
To reach larger scales, human societies have built institutions and other ways of decision making, but the cognitive limit of about 250 contacts remains even in an online society~\cite{gonccalves2011modeling,dunbar2016online}. 
Such results lead to the hypothesis that intelligence and cognitive capabilities are a factor in the ability of AI agent societies to form stable cohesive groups.
This application of concepts and methods from the analysis of human societies to the analysis of AI agent societies is a kind of \textit{AI anthropology}\footnote{In this piece, we adopt language that has a certain degree of anthropomorphising as an abstraction to explain AI agents as in  \url{https://huggingface.co/blog/ethics-soc-7}. We do not mean that AI has a mind of itself but find this language useful to explain complex topics.} that can leverage the insights of one discipline in another.

    \begin{figure*}
        \centering
            \includegraphics[width=0.99\textwidth]{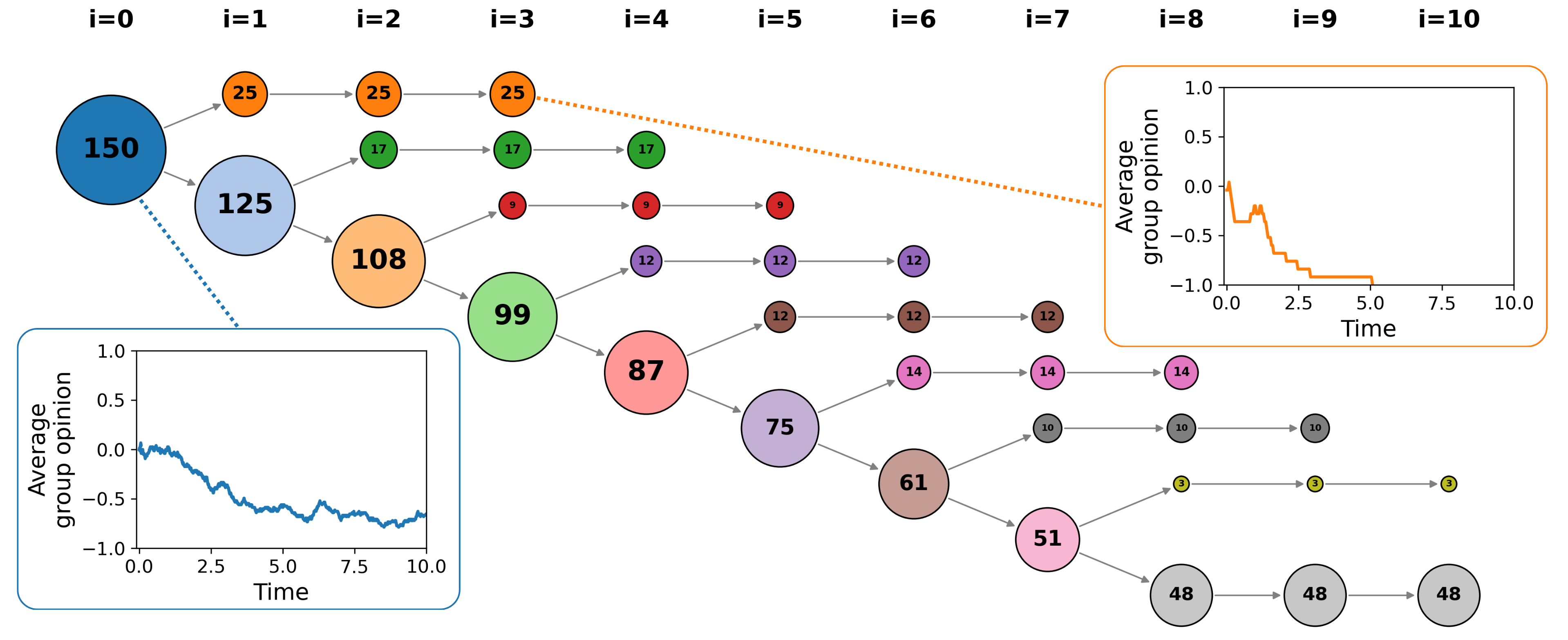}
        \caption{\textbf{Emergence of stable groups.} Simulation of a 150 AI agents powered by Llama 3 70B undergoing iterative opinion formation and splitting. The tree diagram (center) shows how the initial unstable large group progressively fragments into smaller stable cohesive groups through 10 iterations. We consider a group as stable if it does not split after two iterations. Circle sizes represent group population, with numbers indicating agent count. The left graph shows the average opinion over time for the initial 150-agent group. The right graph displays the average opinion for the 25-agent subgroup, which quickly fully coordinates.}     
        \label{fig:plot1}
    \end{figure*}

In this article, we investigate the ability of AI agents to form stable groups, coordinating about decisions for which there is no information supporting one option over another. 
The emergence of coordination is a foundational aspect of social systems, where individual interactions lead to the formation of a unified agreement or shared understanding without the need for a central authority or structure~\cite{dyer2009leadership, baronchelli2018emergence}. 
We develop a framework to test if AI agent societies can form stable groups and up to which scale, and use it to analyze a benchmark of proprietary and open-source models. 
By applying insights from statistical physics, we measure a majority force that determines the possibility of coordination in societies of AI agents. 
This majority force is a function of cognitive capabilities of models and of group size, with coordination not emerging beyond a critical size for a given LLM. Furthermore, we find that AI agents running some of the most capable LLMs are able to coordinate at scales beyond what human groups typically achieve.

\section*{Results}    
\subsection*{Stability of groups of AI agents}
To investigate the coordination abilities of AI agent societies, we perform simulations using agents guided by various LLMs, belonging to the GPT, Claude, and Llama families. 
Simulations run as follows (see Methods for more details). We start with an initial group with $N$ agents. Each agent is assigned an initial opinion randomly chosen from a binary set (e.g., "Opinion A" and "Opinion B"). We then simulate a decision process with binary options, where there is no reason to favor the former over the latter, and vice versa.
At each time step, a single agent is randomly selected to update their opinion. 
The selected agent receives the list of all other agents with their current opinions and is then prompted to choose their new opinion based on this information.
This approach mirrors binary opinion dynamics such as the voter model or Glauber dynamics~\cite{castellano2009statistical}, where agents update their opinions based only on peer interactions. 
However, unlike traditional Agent-Based Models with predefined opinion update rules and equations, here AI agents autonomously decide their opinions based on their LLM. Agents are given some time too coordinate (in our simulations $t=10$, meaning that each agent is updated, on average, $10$ times) and then the simulation is stopped and the state of the system is evaluated. If agents coordinate, i.e., they all agree on the same opinion, the group remains cohesive. On the other hand, if both opinions survive and the AI agents do not all take the same decision, the group splits according to the opinion of the agents. The process is then iterated for the new groups that formed. We stop the evolution of the groups if they remain cohesive for $2$ iterations in a row, and stop the simulation when all groups have stopped this way. This process mimics a group of individuals that undergoes a series of important decisions, each potentially leading to a split. We show in Fig.~\ref{fig:plot1} the result of such a process for Llama 3 70B and initial group size of $N=150$. As can be seen, the group is initially unstable and splitting step after step. However, the groups that result from these splits, when sufficiently small, show stability and do not split further. This shows that while large groups of AI agents are unstable, on smaller scales LLMs are able to coordinate and form cohesive social groups.

To characterize the evolution of the system, we need to focus on the dynamics within single groups. To this extent, we define the average group opinion
            \[
                m=\frac{1}{N}\sum_is_i=\frac{N_+-N_-}{N}.
            \]
Here $s_i$ is the opinion of agent $i$, the first opinion is $s_i=+1$ and the second is $s_i=-1$ (i.e., in favor and against the first option). 
$N_{+}$ and $N_{-}$ are the number of agents supporting the first and second opinion respectively, while $N$ is the total number of agents. 
In these terms we can define the coordination level $C=|m|$ that quantifies the level of agreement among agents. 
Full coordination corresponds to $C=1$, while $C=0$ means that the system is split in two groups of equal size and opposite opinion, thus coordination in absent. Partial coordination is achieved for intermediate values.

    \begin{figure}
        \centering
        \includegraphics[width=0.45\textwidth]{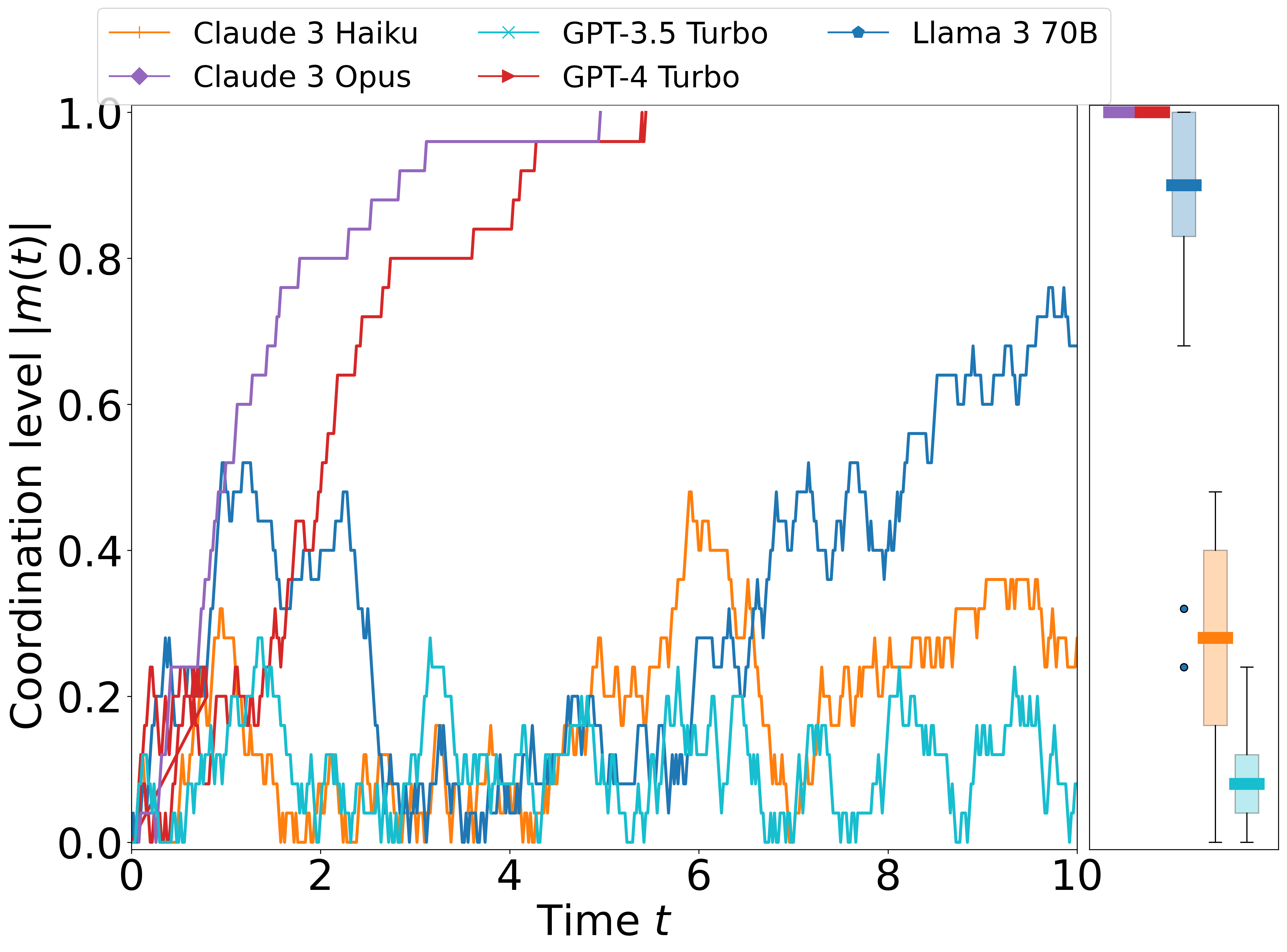}
        \caption{\textbf{Coordination simulations among AI agents.} Evolution of the coordination level over time for five different models and group size $N=50$. The box plots on the right show the final coordination level over $20$ simulations. Some models always fully coordinate, while the others never do so. }
        \label{fig:plot2}
    \end{figure}

We show in Fig.~\ref{fig:plot2} the evolution of the coordination level in societies of $N=50$ agents and five different models, where the boxplot shows $C(t=10)$ over $20$ realizations.
The two most advanced models considered, Claude 3 Opus and GPT-4 Turbo, coordinate in all simulations, which corresponds to $|m|=1$ in the boxplot. 
On the other hand, less advanced models (Claude 3 Haiku and GPT-3.5 Turbo) do not reach coordination in any of the simulations and show an erratic behavior in their coordination level. Finally, Llama 3 70B, a model with intermediate capabilities, shows a clear tendency toward full coordination but does not reach within 10 iterations.

\subsection*{The majority force and its determinants}
\begin{figure*}[t]
\centering
\includegraphics[width=0.99\textwidth]{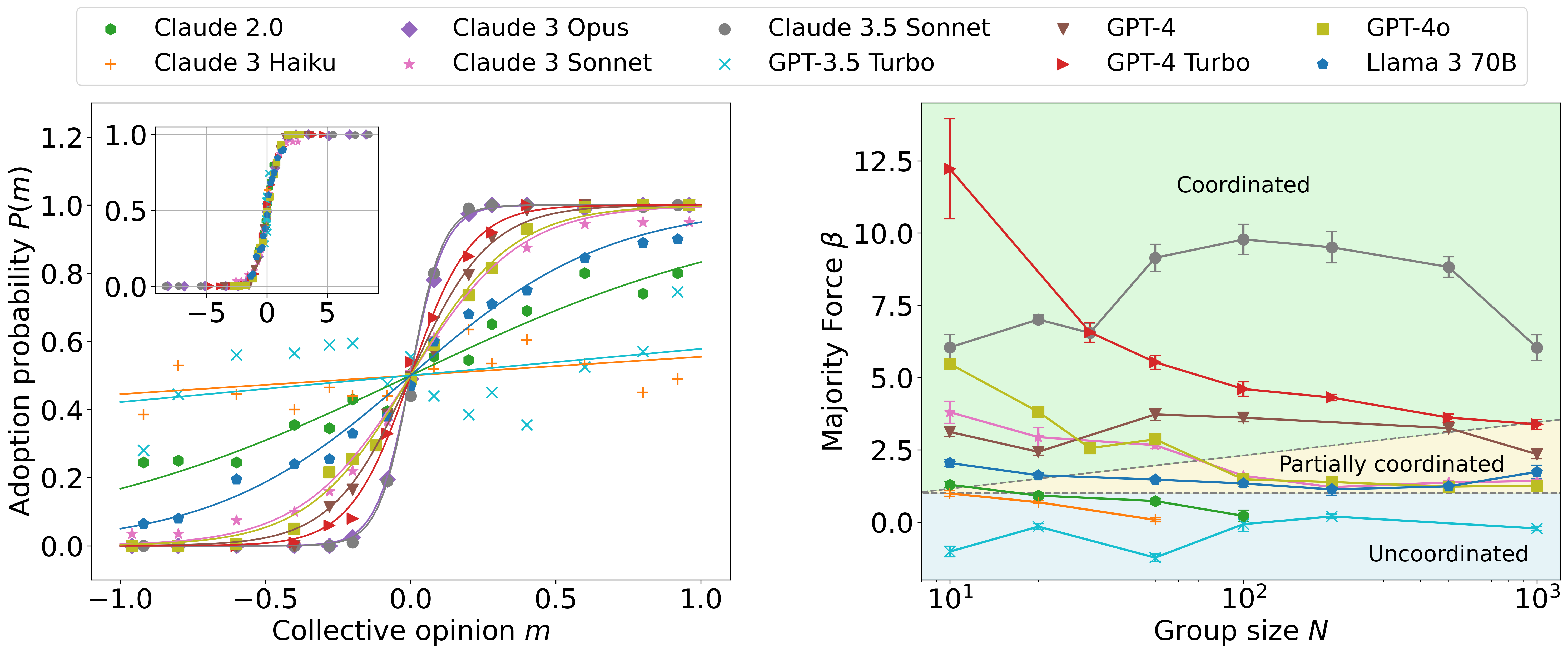}
\caption{\textbf{Adoption probability and majority force.} Left: 
adoption probability $P(m)$ as function of the collective opinion $m$ in a group of size $N=50$. Solid lines are fits of the curve $P(m)=0.5[\tanh(\beta\cdot m)+1]$ to the empirical data. The inset shows rescaled probabilities $P(\tilde{m})$ ($\tilde{m}=\beta m$) and confirms that all LLMs follow the same universal function.
Right: majority force $\beta$ as a function of the group size $N$ for various models. The majority force decreases in larger groups of AI agents. The horizontal dashed line corresponds to $\beta_c=1$, the transition point of the Curie-Weiss model below which no order is present. The growing dot-dashed line indicates $\beta_t(N)$, the crossover point above which fluctuations lead to coordination.}     
\label{fig:plot3}
\end{figure*}

We can get a deeper understanding of the underlying opinion dynamics by looking at the adoption probability $P(m)$, defined as the probability of an agent to be in favor of the first option as function of the average group opinion $m$. 
The left panel of Fig.~\ref{fig:plot3} shows $P(m)$ for ten of the most popular LLMs and $N=50$ agents. 
The adoption probability is an increasing function of $m$ that approaches $1$ when $m=1$ and zero when $m=-1$.
The most advanced models (GPT-4 family, Llama 3 70B, Claude 3 Sonnet and Opus) show a stronger tendency to follow the majority, with more pronounced S-shaped curves.  
On the other hand, less advanced models (GPT-3.5 Turbo, Claude 3 Haiku), have a weaker tendency to follow the majority, with GPT-3.5 Turbo going to some extent against the majority for small values of $m$. 

The dependence of the adoption probability as a function of $m$ approximately follows the function
\begin{equation}
    P(m)=\frac{1}{2}[\tanh(\beta m)+1].
    \label{eq:transition_probability}
\end{equation}
This is a good fit in all cases except GPT-3.5 Turbo.
This can also be seen by looking at the collapse plot showing $P(\tilde{m})$ as a function of the rescaled average opinion $\tilde{m}=m \beta$, shown in the inset of Fig.~\ref{fig:plot3}. All adoption probabilities (except GPT-3.5 Turbo) collapse on the same curve showing a universal behavior. 
The parameter $\beta$, the majority force, gauges the tendency to follow the majority versus randomness in agent's choices.
For $\beta=0$ then $P(m)=1/2$: each agent behaves fully randomly (the new opinion is selected by coin-tossing) and coordination can be reached only by chance. 
This means that the expected time to coordinate grows exponentially with $N$.
For $\beta=\infty$ agents always align with the global majority and coordination is achieved very quickly, on a timescale growing logarithmically with $N$~\cite{castellano2012social}.

Remarkably, the adoption probability in \eqref{eq:transition_probability} is equal to the probability for a spin to be up when the magnetization is $m$ in the Glauber dynamics~\cite{glauber1963time,kochmanski2013curie} for the Curie-Weiss (CW) model, mean-field version of Ising model. The binary opinions can indeed be mapped to $\pm 1$ spin variables and the majority force to an inverse temperature. The equilibrium collective opinion $m^*$ can then be obtained from the self-consistency equation
\begin{equation*}
     m^*=\tanh (\beta m^*)
\end{equation*}
from which the existence of a transition value $\beta_c=1$ can be derived (see Methods). This is a second order phase transition, where order emerges (continuously) for $\beta>\beta_c$~\cite{kochmanski2013curie}. 

Wether or not a group of AI agents can coordinate will then depend on the value of $\beta$, on the corresponding equilibrium value of the collective opinion $m^*$ and on the fluctuations around this value. As we detail in the Methods section, we can identify three different regimes:
\begin{itemize}
        \item \textbf{Absence of Coordination} The majority force $\beta$ is below the critical value $\beta_c=1$. The equilibrium collective opinion is null, so the coordination level $C$ keeps fluctuating close to zero. This means that order is completely absent and the group is constantly in an uncoordinated state;
        \item \textbf{Partial Coordination} The majority force satisfies $\beta_c<\beta<\beta_t$, where $\beta_t\approx (\log N)/2$. The equilibrium collective opinion $m^*$ is larger than zero, but smaller than one. Statistical fluctuations are not strong enough for the system to coordinate. As a consequence $C$ fluctuates around a value greater than $0$ without ever reaching $1$;
        \item \textbf{Coordination} For $\beta>\beta_t$ the equilibrium collective opinion is close to one and fluctuations are enough to coordinate. The coordination level quickly converges to $C=1$.
\end{itemize}
The different behaviors observed in Fig.~\ref{fig:plot2} can then be explained as a direct result of the various LLMs having different majority forces, some of which below the critical value $\beta_c$ or the threshold value $\beta_t$. More details are reported in Methods.

The majority force does not depend only on the type of LLM.
The right panel of Fig.~\ref{fig:plot3} shows the estimated $\beta$ as a function of $N$ for various LLMs. We also show the three different regions in which the $\beta-N$ plane is divided and that corresponds to the three different regimes we discussed above.
\footnote{Given the cost of performing simulations with proprietary models, we analyzed increasing values from $N=10$ and stopped the analysis when $\beta$ reached values clearly below $1$, also excluding Claude 3 Opus in this analysis of $N$ due to its high cost.}
For most models, there is a tendency for sufficiently large $N$: the larger the group, the weaker the majority force $\beta$. It is important to remark that this behavior is connected to the social aspect of the simulation and the prompt provided to the LLMs. No major performance difference is indeed observed in simple counting tasks when varying model or length. Details are reported in the SI. 

In animal (human and non-human) societies, the size of the group plays a crucial role, with a progressive loss of stability of informal groups as the number of individuals increases~\cite{dunbar1992neocortex}. 
Fig.~\ref{fig:plot3} suggests that a similar phenomenon may occur also in AI agent societies, with the ability of agents to coordinate limited by the size of their group. This would explain the behavior observed in Fig.~\ref{fig:plot1}, with stability emerging in small but not in large groups.
        
\subsection*{Critical Group Size}

\begin{figure*}
    \centering
    \includegraphics[width=0.99\textwidth]{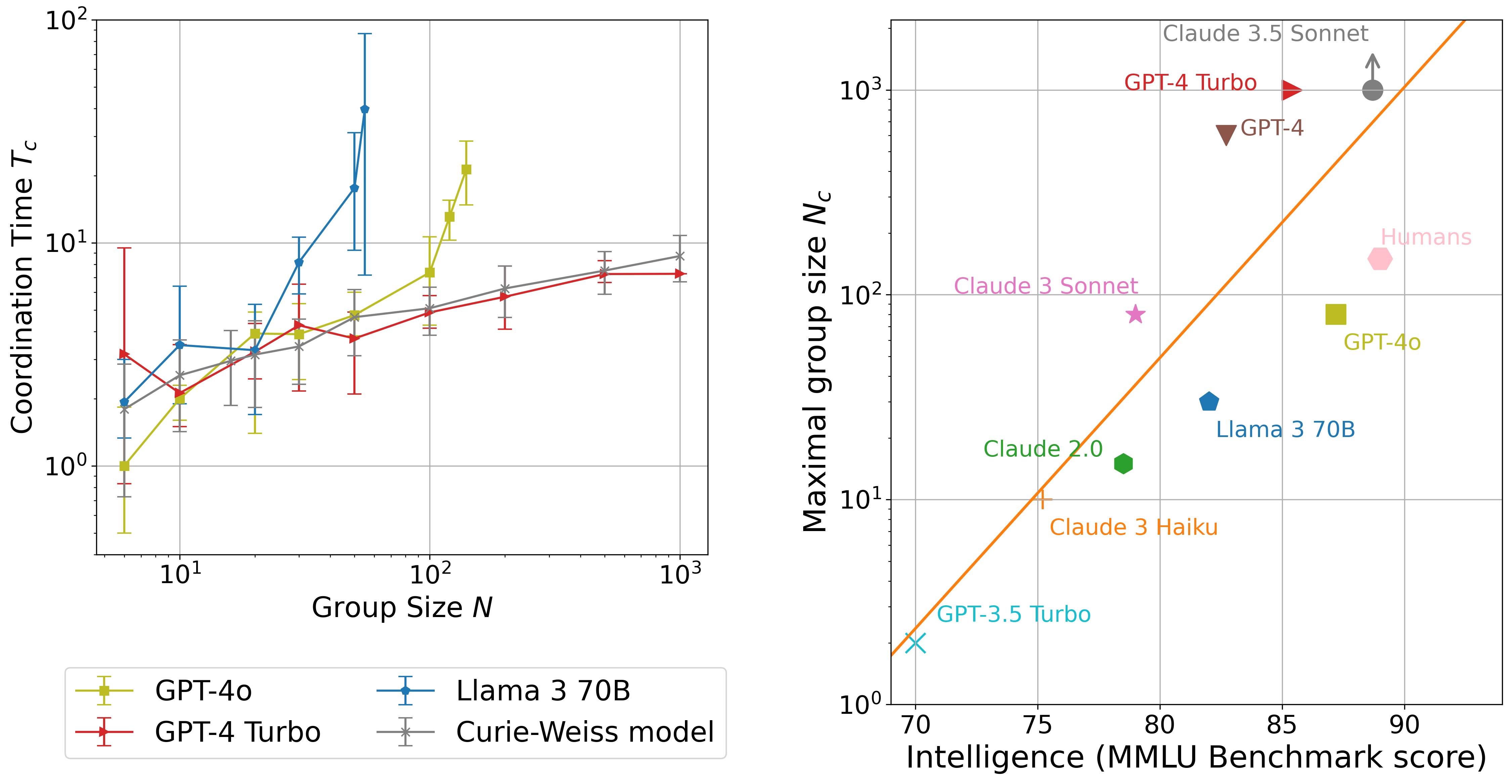}
    \caption{\textbf{Critical Group Size.} 
    Left: Average time to coordinate as a function of the group size for Llama 3 70B, GPT-4 Turbo, GPT-4o and the Curie-Weiss (CW) model with $\beta=3.75$. 
    While both the CW model and GPT-4 Turbo display a slow growth of the coordination time, both Llama 3 70B and GPT-4o exhibit a rapid growth when the group size gets close to the critical group size. 
    Averages are computed over $5$ realizations ($3$ for $N=500$ and only $1$ for $N=1000$ in the case of GPT-4 Turbo) and error bars show the range of values. 
    Right: Critical group size $N_c$ of LLMs after which coordination gets (exponentially) unlikely. 
    For humans, we report Dunbar's number, while for Claude 3.5 Sonnet we can only report a lower bound, since for $N$ up to $1000$ the majority force is well above the threshold value $\beta_t(1000)\approx3.5$. 
    The solid line represents an exponential fit of all points, excluding Claude 3.5 Sonnet and humans. The figure shows that the most capable models exhibits an $N_c$ value much larger than the scale of human informal groups.}     
    \label{fig:plot4}
\end{figure*}
We can characterize more precisely the transition from stable to unstable groups by inspecting, as a function of $N$, the mean time to coordinate starting from random initial opinions. As shown in Methods, this time grows with $N$ differently, depending on whether $\beta$ is larger or smaller than $\beta_t$.
If $\beta>\beta_t$, the coordination time grows logarithmically with $N$~\cite{castellano2012social}, meaning that even a very large group is stable and requires a short time to coordinate. This is the case for GPT-4 Turbo, as shown in Fig.~\ref{fig:plot4}, which perfectly follows the theoretical prediction of the CW model.

When $\beta<\beta_t$ instead the coordination time grows exponentially with $N$, implying in practice that large groups remain incohesive for any reasonable time interval.
The behavior for Llama 3 70B and GPT-4o exhibits both regimes:
when $N$ is small, the coordination time grows slowly, while for larger $N$ it diverges extremely fast. This confirms the presence of a size-induced transition occurring for some critical size $N_c$. We can estimate this value from the size dependence of $\beta$ by setting $\beta(N_c)=\beta_t(N_c)$. For Llama 3 70B $N_c\approx 30$, while for GPT-4o $N_c\approx 80$ (see Fig.~\ref{fig:plot3}). 

The analogy with primates leads to the expectation that $N_c$ depends on the intelligence of the LLMs, just as primates exhibit a growing relationship between neocortex ratio and average group size~\cite{dunbar1992neocortex}.
By studying the values of $\beta$ as a function of $N$ we can determine, for each model, the critical group size $N_c$, where $\beta\approx \beta_t$. We can then compare these values with a measure of models' intelligence. Here we focus on the score in the MMLU (Massive Multitask Language Understanding) benchmark. This is a comprehensive evaluation of a language model's knowledge and reasoning abilities across 57 subjects, ranging from mathematics and law to history and medicine, using multiple-choice questions. Similar results are obtained using also other measures, as detailed in the SI. We show a comparison of MMLU vs $N_c$ on the right panel of Fig.~\ref{fig:plot4}. In the case of Claude 3.5 Sonnet, this provides a lower bound for this quantity as our analysis could not find values $\beta\approx\beta_t$, even for large groups with $N=1000$. We also plot, in the same figure, the critical group size for humans as Dunbar's limit $N_c\approx 200$ \cite{dunbar1992neocortex}.
LLMs display a similar trend as observed in anthropology, with intelligence predicting the limit of coordination size, i.e. the size of groups above which coordination becomes unlikely. 
AI agents show good agreement with an exponentially growing $N_c$ with humans also being very closely aligned to this trend. 
On the other hand, some of the most modern and advanced LLMs reach a super-human coordination capability despite having a human-level MMLU performance. 
This is the case for Claude 3.5 Sonnet, that is well above $\beta_t$ also for $N=1000$, the largest system we considered, and GPT-4 Turbo, for which $N_c\approx 1000$. These values are substantially beyond Dunbar's number and suggest that the ability to coordinate in groups is an emergent property, similarly related, in both humans and in LLMs, to cognitive capabilities. 

\section*{Discussion}
Human societies are characterized by emergent behavior that cannot be understood by studying only individuals in isolation.
The ability to coordinate is one of these emergent group properties that is crucial in the development of languages, social norms, and collective decisions.
For humans, Dunbar's number $N_c\approx 200$ sets the maximal number of personal relations we can maintain and thus also the maximal group size in which consensus and coordination, intended as the spontaneous emergence of common social decisions, can exist.
Studying humans and other primates, researchers have identified a power-law scaling connecting the neocortex ratio to the average group size \cite{dunbar1992neocortex}, thus proving the link between cognitive capabilities and the development of large societies. 

LLMs are attracting a growing interest in the social sciences for their ability to mimic humans, both at the individual and, as recent studies suggest, at the group level.
Like humans, AI agents show emergent group properties that are not directly coded in their training process.
We argue that the ability to coordinate and form cohesive groups is one of these properties, with AI agents showing striking similarities with primates including humans.
As a first result, we show that all the most advanced LLMs are characterized by a majority-following trend described by a universal function with a single parameter $\beta$, the majority force.
Remarkably, this function depends on the specific model but its analytic form is the same describing magnetic spin systems.
Different models typically have different $\beta$, with the less sophisticated ones showing smaller $\beta$, i.e., a weaker majority force and thus a more erratic behavior that prevents coordination. 
The majority force depends not only on the LLM, but also on the group size: it tends to be larger in smaller groups.
This evidence and the analogy with the Curie-Weiss model allowed us to compute ``Dunbar's number'' of each LLM, a size threshold above which a society of agents driven by that LLM is too large to be cohesive and tends to split.
While most models show human-like scaling with cognitive capabilities, some of the most sophisticated LLMs are capable coordinate in groups in the thousands, going beyond what human groups can do without rules and institutions.
           
These results are important due to the relevance of collective behavior and coordination in social contexts.
More research is needed to understand other conditions that lead to the emergence of coordination in AI agent societies, especially for other types of opinion dynamics with incentives, unequal information access, or other kinds of diversity.
The ability of LLMs to coordinate can be beneficial, for example when aiming at coordinating group activities of LLMs where no clear preference for any of the options exists.
When there is no information to guide how to behave, AI agent societies could spontaneously reach their own social norms, making their behavior predictable by other agents, despite the absence of an intrinsically preferrable choice. 
However, this also poses threats, as these norms might not be aligned with human values or in situations where coordination threatens the integrity of a system, such as the case of flash crashes among trading bots.
Future research with this AI Anthropology perspective is needed to understand better how this kind of coordination can happen in practical scenarios beyond the idealized situation we studied here.
To work towards responsible and trustworthy AI, we need to investigate systemic risks stemming from the collective behavior of AI agents, where coordination as we showed here is a first example.

\section*{Methods}
\subsection*{Opinion dynamics simulations}
We implement an opinion dynamics process for binary opinions with memoryless AI agents. 
\begin{enumerate}
    \item At each step an agent is randomly selected and time is incremented by $dt=1/N$;
    \item the agent is given the full list of all agents in the system, each identified by a random name, and the opinions they support. Note that the agent's own opinion is not included in this prompt;
    \item the selected agent is asked to reply with the opinion they want to support and their opinion is updated correspondingly; 
    \item the process is then iterated until coordination is reached or until the time reaches the maximum preset limit.
\end{enumerate}
Note that in one time unit, $t\to t+1$, $N$ updates are performed, so that, on average, each LLMs is selected once. 
In all simulations the initial collective opinion $m$ is set to zero, i.e., initially the same number of agents supports each of the two opinions. Following the framework introduced in~\cite{de2023emergence}, steps 2-3 are performed using this prompt:\\
\centerline{\parbox{8cm}{\textit{
                            Below you can see the list of all your friends together with the opinion they support.\\
                            You must reply with the opinion you want to support.\\
                            The opinion must be reported between square brackets.\\
                            X7v A\\
                            keY B\\
                            91c B\\
                            gew A\\
                            4lO B\\
                            ...\\
                            ...\\
                            Reply only with the opinion you want to support, between square brackets.
}}}

Here $A$ and $B$ are the opinion names. Most LLMs exhibit an opinion bias, with a tendency to prefer one opinion name over the other. 
This bias is particularly strong when the names have an intrinsic meaning, like for instance ``Yes'' and ``No''. In such a case LLMs display a strong preference toward the more ``positive'' opinion, preferring ``Yes'' over ``No'' (see SI). 
For this reason it is important to use letters or random combinations of them as opinion names. 
Even doing so, small biases are typically present. 
However, they are much weaker than for meaningful names and they can be easily removed by performing a random shuffling of opinion names at each iteration. 
For instance at $t=0$ the first opinion may be called $k$ and the second $z$, while at $t=dt$ these names are swapped with probability $0.5$, meaning that the first opinion is now called $z$ and the second $k$, while keeping the reference unchanged in our analysis. 
In all our simulations we used the opinion names $k$ and $z$; we tested that using different opinion names does not cause any significant difference. We also tested the robustness to prompts variation observing an overall stability, with the most advanced models showing little to no variability and less advanced models presenting some variations in the behaviour. Details on the robustness tests are reported in the SI.

\subsection*{Details on the LLMs}
Table~\ref{tab:list_models} reports the model version of all the LLMs considered.
\begin{table}[t]
    \centering
    \begin{tabular}{|l|l|}
    \hline
    \textbf{Model Name} & \textbf{Model Version}               \\ \hline
    Claude 3.5 Sonnet               & claude-3-5-sonnet-20240620        \\ \hline
    Claude 3 Haiku                  & claude-3-haiku-20240307           \\ \hline
    Claude 3 Opus                   & claude-3-opus-20240229            \\ \hline
    Claude 3 Sonnet                 & claude-3-sonnet-20240229          \\ \hline
    Claude 2.0                      & claude-2.0                        \\ \hline
    GPT-3.5 Turbo                   & gpt-3.5-turbo-1106                \\ \hline
    GPT-4                           & gpt-4-0613                        \\ \hline
    GPT-4o                          & gpt-4o-2024-05-13                 \\ \hline
    GPT-4 Turbo                     & gpt-4-turbo-2024-04-09            \\ \hline
    Llama 3 70B                     & meta-llama-3-70b-instruct         \\ \hline
    \end{tabular}
    \caption{Specific Model Versions we used in our simulations}
    \label{tab:list_models}
\end{table}

LLMs are characterized by a temperature parameter $T$ determining the variance in the sampling of tokens when they respond to prompts, such that higher temperatures sample with more variance to the same prompt.
In all the simulations reported in the main text we used $T=0.2$. 
As detailed in the SI, there are no relevant changes when different values of $T$ are used, but a low temperature ensures reliability in the output format.

\subsection*{Curie-Weiss Model}
The Curie-Weiss (CW) model is arguably the simplest model of a ferromagnet. 
It describes a system of $N$ spins that can only have two states, either $s_i=+1$ (up) or $s_i=-1$ (down), coupled with ferromagnetic interactions, i.e., favoring mutual alignment.
Each spin interacts with all the others, as the interaction pattern is a fully connected network. 
The CW model is the mean-field limit of the well-known Ising model. 
The magnetization $m$, defined as the average of the spin values $m=\langle s_i\rangle$ is the equivalent of the average group opinion. 
The connection between our LLM based opinion simulations and the CW model derives from the transition probability defined by~\eqref{eq:transition_probability}. 
This expression is indeed equal to the transition probability of Glauber dynamics~\cite{glauber1963time}, which allows to simulate the CW model by means of a Markov chain Monte Carlo approach.

The equilibrium value of the magnetization in the CW model, which is reached from any initial configuration,  is given by the self-consistency equation 
\begin{equation}
     m^*=\tanh(\beta m^*).
     \label{eq:met_self_consistent}
\end{equation}
The solution of this equation depends on the value of the inverse temperature $\beta$ (which corresponds, in the opinion dynamics framework, to the majority force). 
For $\beta<1$ the only solution is $m^*=0$, while for $\beta>1$, $m^*=0$ is an unstable solution, while two new solutions $\pm m^*(\beta)$ appear. 
The value $\beta_c=1$ is the critical point of a second order phase transition. This means that as soon as $\beta>1$, $m^*$ grows gradually with $\beta$, tending to $m^*=1$ for large values of $\beta$, approximately following the equation (see SI)
\[
    m^* = \sqrt{3(\beta-1)}.
\]

\subsection*{Critical group size}
For $\beta>\beta_c$ the equilibrium value of the magnetization is $|m^*|>0$, but this does not mean that coordination ($|m|=1$) is reached.
The magnetization fluctuates over time around $m^*$, the amplitude of fluctuations decreasing with the number of agents $N$.
Full coordination is reached when a fluctuation leads from $m^*$ to $|m|=1$.
Therefore the ability to coordinate crucially depends on the interplay between the value of $m^*$ and $N$. 
If $N$ is not too "large" (in a sense specified below) normal fluctuations are sufficient to rapidly lead to a fully ordered configuration with $|m|=1$. In this case the time to coordinate grows logarithmically with $N$.
On the other hand, if $N$ is too "large", coordination can be achieved only when a large and extremely rare fluctuation around $m^*$ occurs. The average time for this rare fluctuation grows exponentially with $N$
and hence coordination becomes practically unreachable in reasonable times.
Since the equilibrium collective opinion $m^*$ is determined by $\beta$, this implies that a crossover value $\beta_t(N)$ separates the two regimes.

To get a more precise picture we need to study the stochastic differential equation governing the out of equilibrium evolution of $m$. If we are interested in the behavior around the equilibrium magnetization $m^*\approx1$, we can approximate it with an Ornstein–Uhlenbeck process
\[
    dm=-(m-m^*)dt+\sqrt{2D}dW(t),
\]
with
\[
    D\approx \frac{2}{N}\text{e}^{-2\beta}
\]
More detailed computations are reported in SI.
The collective opinion will move randomly around $m^*$ with fluctuations of the order of the variance of $m$, that satisfies
\begin{equation}
    \text{Var}[m]\approx D \approx \frac{2}{N}\text{e}^{-2\beta}
    \label{eq:met_variance_m}
\end{equation}

This estimate of the typical fluctuations allows to determine the crossover value $\beta_t$ separating a regime
($\beta > \beta_t$) where fluctuations of the size of the variance lead to coordination, from a regime ($\beta < \beta_t$) where a large rare fluctuation is instead necessary. As shown in the SI it holds
\begin{equation}
\beta_t\approx\frac{1}{2}\log{N}.
\label{eq:betatvsN}
\end{equation}
\eqref{eq:betatvsN} is reported in Fig.~\ref{fig:plot3}. As expected $\beta_t>\beta_c=1$ for $N$ larger than a few units. Note that for what concerns the time to coordinate there is little difference between systems with $\beta_c<\beta<\beta_t$ and "disordered" ones with $\beta<\beta_c$. In the former case fluctuations occur around a finite value of the magnetization, in the latter around $m^*=0$. But in both cases coordination can be achieved only because of an anomalously large rare fluctuation, requiring exponentially large times to develop. Finally, we can estimate the critical group size $N_c$ of AI agents as the size for which the majority force reaches $\beta_t$
\[
    \beta(N_c)=\frac{1}{2}\log{N_c}.
\]
 
\section*{Supplementary information}
Additional analysis and robustness tests are reported in the Supplementary Information

\section*{Declarations}
\subsection*{Funding}
This project was supported by OpenAI with free API credits under its research programme.
\subsection*{Code availability}
All code is publicly available at \url{https://github.com/giordano-demarzo/LLMs-Opinion-Dynamics}

\subsection*{Acknowledgments}
We are grateful to Profs. Andrea Baronchelli, Márton Pósfai and Viola Priesemann for interesting discussions. 

\bibliographystyle{plain}

\begin{thebibliography}{10}

\bibitem{aher2023using}
Gati~V Aher, Rosa~I Arriaga, and Adam~Tauman Kalai.
\newblock Using large language models to simulate multiple humans and replicate
  human subject studies.
\newblock In {\em International Conference on Machine Learning}, pages
  337--371. PMLR, 2023.

\bibitem{argyle2023out}
Lisa~P Argyle, Ethan~C Busby, Nancy Fulda, Joshua~R Gubler, Christopher
  Rytting, and David Wingate.
\newblock Out of one, many: Using language models to simulate human samples.
\newblock {\em Political Analysis}, 31(3):337--351, 2023.

\bibitem{aroyehun2023leia}
Segun~Taofeek Aroyehun, Lukas Malik, Hannah Metzler, Nikolas Haimerl, Anna
  Di~Natale, and David Garcia.
\newblock Leia: Linguistic embeddings for the identification of affect.
\newblock {\em EPJ Data Science}, 12(1):52, 2023.

\bibitem{bail2024can}
Christopher~A Bail.
\newblock Can generative ai improve social science?
\newblock {\em Proceedings of the National Academy of Sciences},
  121(21):e2314021121, 2024.

\bibitem{baronchelli2018emergence}
Andrea Baronchelli.
\newblock The emergence of consensus: a primer.
\newblock {\em Royal Society open science}, 5(2):172189, 2018.

\bibitem{binz2023using}
Marcel Binz and Eric Schulz.
\newblock Using cognitive psychology to understand gpt-3.
\newblock {\em Proceedings of the National Academy of Sciences},
  120(6):e2218523120, 2023.

\bibitem{boiko2023autonomous}
Daniil~A Boiko, Robert MacKnight, Ben Kline, and Gabe Gomes.
\newblock Autonomous chemical research with large language models.
\newblock {\em Nature}, 624(7992):570--578, 2023.

\bibitem{casari2018group}
Marco Casari and Claudio Tagliapietra.
\newblock Group size in social-ecological systems.
\newblock {\em Proceedings of the National Academy of Sciences},
  115(11):2728--2733, 2018.

\bibitem{castellano2012social}
Claudio Castellano.
\newblock Social influence and the dynamics of opinions: The approach of
  statistical physics.
\newblock {\em Managerial and Decision Economics}, 33(5-6):311--321, 2012.

\bibitem{castellano2009statistical}
Claudio Castellano, Santo Fortunato, and Vittorio Loreto.
\newblock Statistical physics of social dynamics.
\newblock {\em Reviews of modern physics}, 81(2):591--646, 2009.

\bibitem{cau2025language}
Erica Cau, Valentina Pansanella, Dino Pedreschi, and Giulio Rossetti.
\newblock Language-driven opinion dynamics in agent-based simulations with
  llms.
\newblock {\em arXiv preprint arXiv:2502.19098}, 2025.

\bibitem{chang2024llms}
Serina Chang, Alicja Chaszczewicz, Emma Wang, Maya Josifovska, Emma Pierson,
  and Jure Leskovec.
\newblock Llms generate structurally realistic social networks but overestimate
  political homophily.
\newblock {\em arXiv preprint arXiv:2408.16629}, 2024.

\bibitem{chang2023booookscore}
Yapei Chang, Kyle Lo, Tanya Goyal, and Mohit Iyyer.
\newblock Booookscore: A systematic exploration of book-length summarization in
  the era of llms.
\newblock {\em arXiv preprint arXiv:2310.00785}, 2023.

\bibitem{chuang2023simulating}
Yun-Shiuan Chuang, Agam Goyal, Nikunj Harlalka, Siddharth Suresh, Robert
  Hawkins, Sijia Yang, Dhavan Shah, Junjie Hu, and Timothy~T Rogers.
\newblock Simulating opinion dynamics with networks of llm-based agents.
\newblock {\em arXiv preprint arXiv:2311.09618}, 2023.

\bibitem{couzin2005effective}
Iain~D Couzin, Jens Krause, Nigel~R Franks, and Simon~A Levin.
\newblock Effective leadership and decision-making in animal groups on the
  move.
\newblock {\em Nature}, 433(7025):513--516, 2005.

\bibitem{de2023emergence}
Giordano De~Marzo, Luciano Pietronero, and David Garcia.
\newblock Emergence of scale-free networks in social interactions among large
  language models.
\newblock {\em arXiv preprint arXiv:2312.06619}, 2023.

\bibitem{dentella2023systematic}
Vittoria Dentella, Fritz G{\"u}nther, and Evelina Leivada.
\newblock Systematic testing of three language models reveals low language
  accuracy, absence of response stability, and a yes-response bias.
\newblock {\em Proceedings of the National Academy of Sciences},
  120(51):e2309583120, 2023.

\bibitem{dunbar1992neocortex}
Robin~IM Dunbar.
\newblock Neocortex size as a constraint on group size in primates.
\newblock {\em Journal of human evolution}, 22(6):469--493, 1992.

\bibitem{dunbar1998social}
Robin~IM Dunbar.
\newblock The social brain hypothesis.
\newblock {\em Evolutionary Anthropology: Issues, News, and Reviews: Issues,
  News, and Reviews}, 6(5):178--190, 1998.

\bibitem{dunbar1999culture}
Robin~IM Dunbar.
\newblock Culture, honesty and the freerider problem.
\newblock {\em The evolution of culture}, pages 194--213, 1999.

\bibitem{dunbar2016online}
Robin~IM Dunbar.
\newblock Do online social media cut through the constraints that limit the
  size of offline social networks?
\newblock {\em Royal Society Open Science}, 3(1):150292, 2016.

\bibitem{dyer2009leadership}
John~RG Dyer, Anders Johansson, Dirk Helbing, Iain~D Couzin, and Jens Krause.
\newblock Leadership, consensus decision making and collective behaviour in
  humans.
\newblock {\em Philosophical Transactions of the Royal Society B: Biological
  Sciences}, 364(1518):781--789, 2009.

\bibitem{glauber1963time}
Roy~J Glauber.
\newblock Time-dependent statistics of the ising model.
\newblock {\em Journal of mathematical physics}, 4(2):294--307, 1963.

\bibitem{gonccalves2011modeling}
Bruno Gon{\c{c}}alves, Nicola Perra, and Alessandro Vespignani.
\newblock Modeling users' activity on twitter networks: Validation of dunbar's
  number.
\newblock {\em PloS one}, 6(8):e22656, 2011.

\bibitem{grossmann2023ai}
Igor Grossmann, Matthew Feinberg, Dawn~C Parker, Nicholas~A Christakis,
  Philip~E Tetlock, and William~A Cunningham.
\newblock Ai and the transformation of social science research.
\newblock {\em Science}, 380(6650):1108--1109, 2023.

\bibitem{guo2024embodied}
Xudong Guo, Kaixuan Huang, Jiale Liu, Wenhui Fan, Natalia V{\'e}lez, Qingyun
  Wu, Huazheng Wang, Thomas~L Griffiths, and Mengdi Wang.
\newblock Embodied llm agents learn to cooperate in organized teams.
\newblock {\em arXiv preprint arXiv:2403.12482}, 2024.

\bibitem{jiang2023llm}
Dongfu Jiang, Xiang Ren, and Bill~Yuchen Lin.
\newblock Llm-blender: Ensembling large language models with pairwise ranking
  and generative fusion.
\newblock {\em arXiv preprint arXiv:2306.02561}, 2023.

\bibitem{johnson2013abrupt}
Neil Johnson, Guannan Zhao, Eric Hunsader, Hong Qi, Nicholas Johnson, Jing
  Meng, and Brian Tivnan.
\newblock Abrupt rise of new machine ecology beyond human response time.
\newblock {\em Scientific reports}, 3(1):2627, 2013.

\bibitem{kochmanski2013curie}
Martin Kochma{\'n}ski, Tadeusz Paszkiewicz, and S{\l}awomir Wolski.
\newblock Curie--weiss magnet—a simple model of phase transition.
\newblock {\em European Journal of Physics}, 34(6):1555, 2013.

\bibitem{liu2023dynamic}
Zijun Liu, Yanzhe Zhang, Peng Li, Yang Liu, and Diyi Yang.
\newblock Dynamic llm-agent network: An llm-agent collaboration framework with
  agent team optimization.
\newblock {\em arXiv preprint arXiv:2310.02170}, 2023.

\bibitem{lu2024generative}
Yikang Lu, Alberto Aleta, Chunpeng Du, Lei Shi, and Yamir Moreno.
\newblock Generative agent-based models for complex systems research: a review.
\newblock {\em arXiv preprint arXiv:2408.09175}, 2024.

\bibitem{miah2024multimodal}
Md~Saef~Ullah Miah, Md~Mohsin Kabir, Talha~Bin Sarwar, Mejdl Safran, Sultan
  Alfarhood, and MF~Mridha.
\newblock A multimodal approach to cross-lingual sentiment analysis with
  ensemble of transformer and llm.
\newblock {\em Scientific Reports}, 14(1):9603, 2024.

\bibitem{papachristou2024network}
Marios Papachristou and Yuan Yuan.
\newblock Network formation and dynamics among multi-llms.
\newblock {\em arXiv preprint arXiv:2402.10659}, 2024.

\bibitem{park2023generative}
Joon~Sung Park, Joseph O'Brien, Carrie~Jun Cai, Meredith~Ringel Morris, Percy
  Liang, and Michael~S Bernstein.
\newblock Generative agents: Interactive simulacra of human behavior.
\newblock In {\em Proceedings of the 36th annual acm symposium on user
  interface software and technology}, pages 1--22, 2023.

\bibitem{park2022social}
Joon~Sung Park, Lindsay Popowski, Carrie Cai, Meredith~Ringel Morris, Percy
  Liang, and Michael~S Bernstein.
\newblock Social simulacra: Creating populated prototypes for social computing
  systems.
\newblock In {\em Proceedings of the 35th Annual ACM Symposium on User
  Interface Software and Technology}, pages 1--18, 2022.

\bibitem{pellert2023ai}
Max Pellert, Clemens~M Lechner, Claudia Wagner, Beatrice Rammstedt, and Markus
  Strohmaier.
\newblock Ai psychometrics: Assessing the psychological profiles of large
  language models through psychometric inventories.
\newblock {\em Perspectives on Psychological Science}, page 17456916231214460,
  2023.

\bibitem{romera2024mathematical}
Bernardino Romera-Paredes, Mohammadamin Barekatain, Alexander Novikov, Matej
  Balog, M~Pawan Kumar, Emilien Dupont, Francisco~JR Ruiz, Jordan~S Ellenberg,
  Pengming Wang, Omar Fawzi, et~al.
\newblock Mathematical discoveries from program search with large language
  models.
\newblock {\em Nature}, 625(7995):468--475, 2024.

\bibitem{rossetti2024social}
Giulio Rossetti, Massimo Stella, R{\'e}my Cazabet, Katherine Abramski, Erica
  Cau, Salvatore Citraro, Andrea Failla, Riccardo Improta, Virginia Morini, and
  Valentina Pansanella.
\newblock Y social: an llm-powered social media digital twin.
\newblock {\em arXiv preprint arXiv:2408.00818}, 2024.

\bibitem{shen2024hugginggpt}
Yongliang Shen, Kaitao Song, Xu~Tan, Dongsheng Li, Weiming Lu, and Yueting
  Zhuang.
\newblock Hugginggpt: Solving ai tasks with chatgpt and its friends in hugging
  face.
\newblock {\em Advances in Neural Information Processing Systems}, 36, 2024.

\bibitem{strachan2024testing}
James~WA Strachan, Dalila Albergo, Giulia Borghini, Oriana Pansardi, Eugenio
  Scaliti, Saurabh Gupta, Krati Saxena, Alessandro Rufo, Stefano Panzeri, Guido
  Manzi, et~al.
\newblock Testing theory of mind in large language models and humans.
\newblock {\em Nature Human Behaviour}, pages 1--11, 2024.

\bibitem{tornberg2023simulating}
Petter T{\"o}rnberg, Diliara Valeeva, Justus Uitermark, and Christopher Bail.
\newblock Simulating social media using large language models to evaluate
  alternative news feed algorithms.
\newblock {\em arXiv preprint arXiv:2310.05984}, 2023.

\bibitem{wu2023autogen}
Qingyun Wu, Gagan Bansal, Jieyu Zhang, Yiran Wu, Shaokun Zhang, Erkang Zhu,
  Beibin Li, Li~Jiang, Xiaoyun Zhang, and Chi Wang.
\newblock Autogen: Enabling next-gen llm applications via multi-agent
  conversation framework.
\newblock {\em arXiv preprint arXiv:2308.08155}, 2023.

\bibitem{zheng2023gpt}
Shen Zheng, Yuyu Zhang, Yijie Zhu, Chenguang Xi, Pengyang Gao, Xun Zhou, and
  Kevin Chen-Chuan Chang.
\newblock Gpt-fathom: Benchmarking large language models to decipher the
  evolutionary path towards gpt-4 and beyond.
\newblock {\em arXiv preprint arXiv:2309.16583}, 2023.

\bibitem{zhou2005discrete}
W-X Zhou, Didier Sornette, Russell~A Hill, and Robin~IM Dunbar.
\newblock Discrete hierarchical organization of social group sizes.
\newblock {\em Proceedings of the Royal Society B: Biological Sciences},
  272(1561):439--444, 2005.

\end{thebibliography}

\newpage
\section*{Supplementary Information}
\subsection*{Bias removal}
            \begin{figure*}
			\centering
                \includegraphics[width=0.75\textwidth]{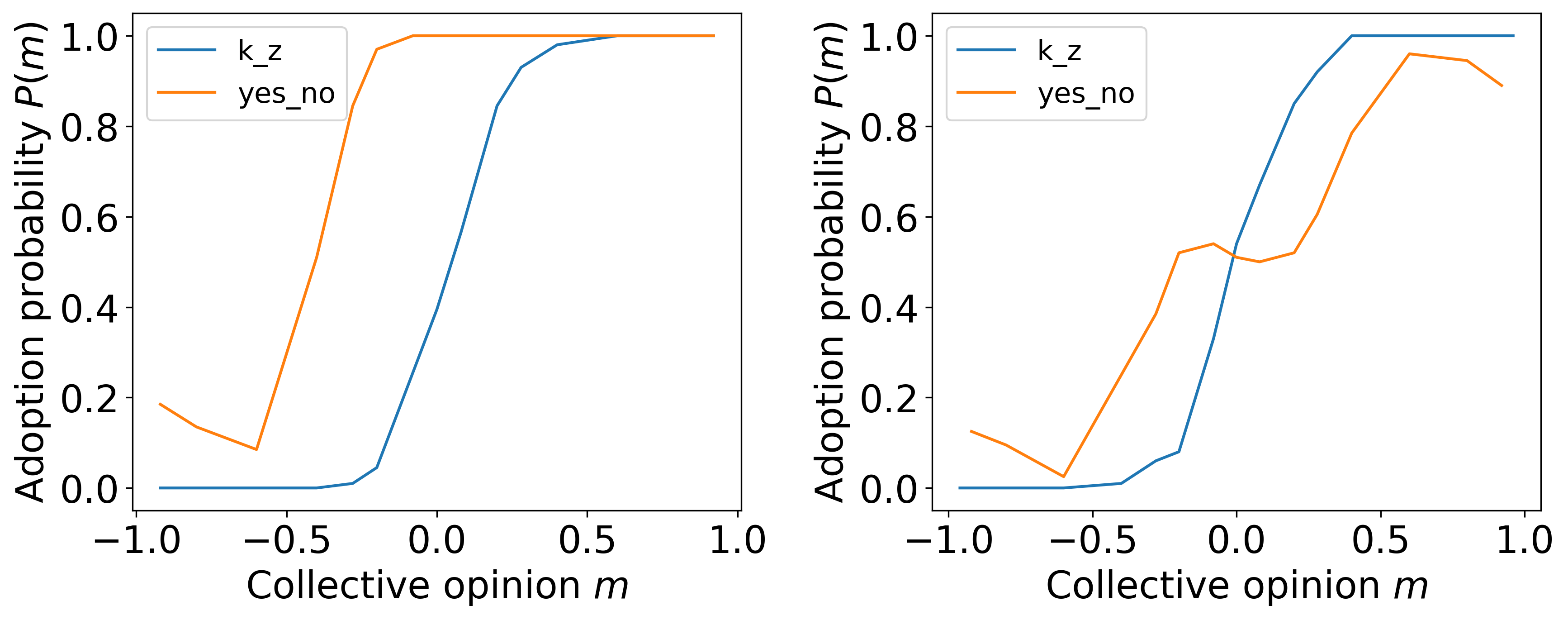}
			\caption{\textbf{Bias and shuffling.} Left panel: Adoption probability $P(m)$ for $N=50$, $T=0.2$ without the shuffling procedure. The set of opinions ``yes, no'' shows a remarkable bias, with LLMs preferring the opinion ``yes''. Right panel: As for the left panel, but with the shuffling procedure in place. This produces a $\tanh$ like adoption probability for the opinions ``k, z'', while for ``yes, no'' the bias is too strong and the shuffling results in a non monotonic adoption probability not converging to $1$ ($0$) when the collective opinion is $+1$ ($-1$).}     
			\label{fig:plot1_SI}
	\end{figure*}
        As mentioned in the main text, in order to remove opinion biases we have to shuffle the opinion names at each iteration. However, this only works if the initial bias is not too strong. We show in Fig.~\ref{fig:plot1_SI} the adoption probability with and without shuffling for two opinion names combinations: ``yes, no'' and ``k, z''. Clearly the former has a very strong bias toward ``yes'' and the shuffling procedure results in an adoption probability different from a $\tanh$ function. On the other hand, ``k, z'' present a very mild bias and the shuffling procedure allows such a bias to be removed without altering the shape of the adoption probability.
        
    \subsection*{Role of opinion names}
        \begin{figure*}
				\centering
				\includegraphics[width=0.95\textwidth]{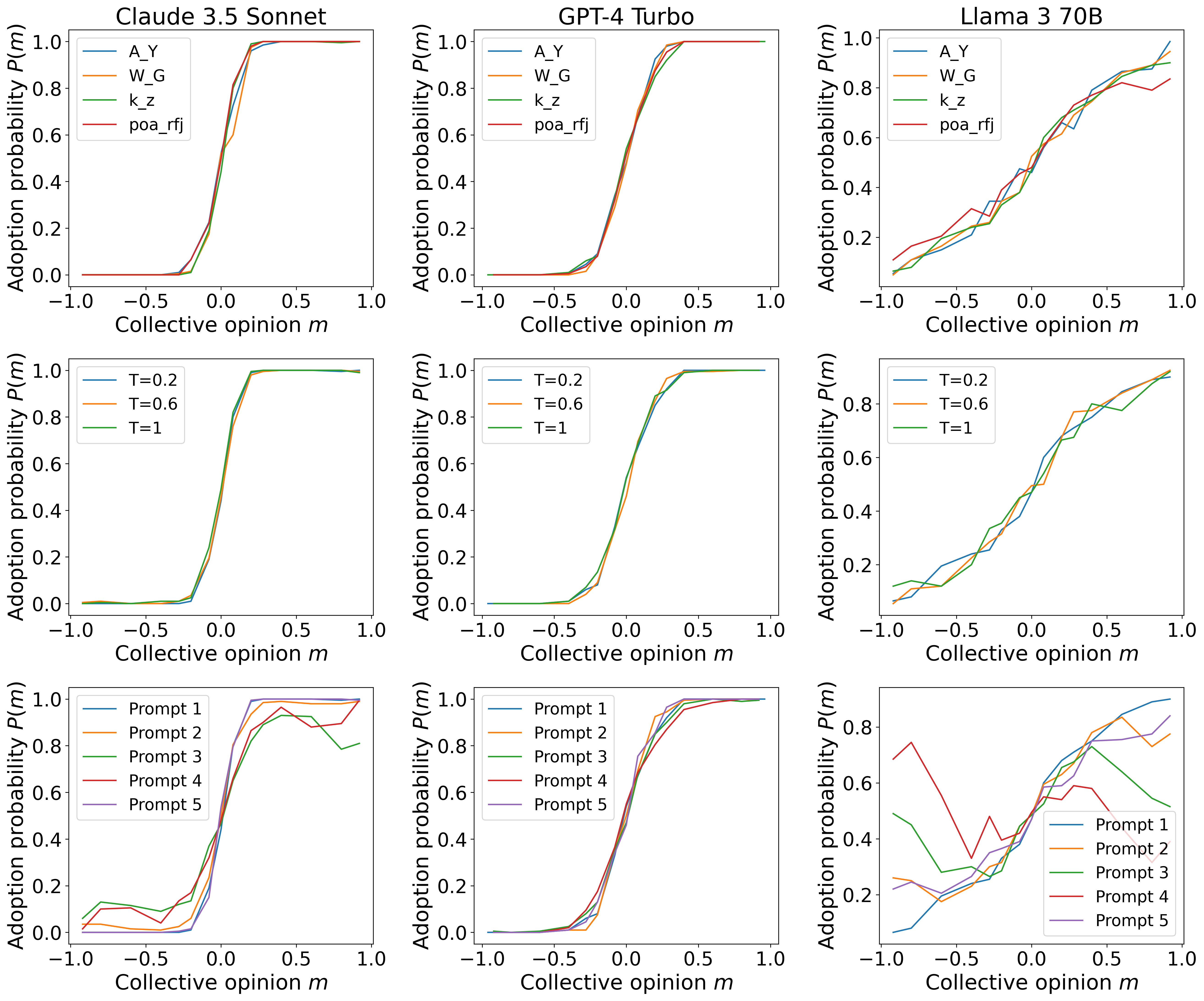}
				\caption{\textbf{Role of opinion names and temperature.} Top plots: Adoption probability $P(m)$ for $N=50$, $T=0.2$ and four different pairs of opinion names. The probability is computed over $200$ simulations. Bottom plots: Adoption probability $P(m)$ for $N=50$, opinion names k, z and three different temperatures. The probability is computed over $200$ simulations.}     
				\label{fig:plot2_SI}
	\end{figure*}
 
        In order to test the stability of the results we investigate the shape of the adoption probability when considering different opinion names. We report in Fig.~\ref{fig:plot2_SI} (top row) the results of this procedure for four possible pairs of opinion names and three different LLMs, representative of the three families of models studied in this work. It turns out that only in the case of Llama there is a difference and only for one of the name pairs considered. In any case, the functional form of the adoption probability is always the same and therefore the general picture is not affected by these minor variations. The most advanced models, GPT-4 Turbo and Claude 3.5 Sonnet, show no differences at all, indicating that, as the model becomes more capable, biases and differences due to the opinion names disappear.
        
    \subsection*{Role of Model temperature}
        Another aspect we tested is the effect of the model temperature $T$. This parameter sets the level of creativity or randomness of the LLM. For $T=0$ the model behaves deterministically, always producing in output the token (word) with the highest probability. Instead, when $T>0$, randomness starts to play a role and also other tokens can be observed in the output. As shown in Fig.~\ref{fig:plot2_SI} (central row), we considered three different temperatures $T=0.2, 0.6, 1.0$ observing no substantial difference in the adoption probability.
   \subsection*{Prompt robustness}
   		As another robustness text we compute the adoption probability for five different prompts. All prompts contain more or less the same request, but it is formulated in different ways. We also explored the role played by explicitly mentioning ``friends'' with respect to just ``people''. The five prompts we experimented with are 
   		\begin{itemize}
   			\item \textbf{Prompt 1}\\
   			\textit{Below you can see the list of all your friends together with the opinion they support.\\
   					You must reply with the opinion you want to support.\\
                    The opinion must be reported between square brackets.}
            \item \textbf{Prompt 2}\\
   			\textit{Below you can see a list of people together with the opinion they support.\\
       				You must reply with the opinion you want to support.\\
       				The opinion must be reported between square brackets.}
            \item \textbf{Prompt 3}\\
   			\textit{The list below contains people along with the opinions they endorse.\\
   					Please respond with the opinion you'd like to support.\\
        			Be sure to enclose the opinion in square brackets.}    
            \item \textbf{Prompt 4}\\
   			\textit{You recently subscribed to a social network.\\ 
       				Below you can see the list of all your friends together with the group they joined on the social network.\\
       				You must reply with the name of the group you want to join.\\
      				The name of the group must be reported between square brackets.}
            \item \textbf{Prompt 5}\\
   			\textit{You recently subscribed to a social network.\\
        			Below you can see the list of all your friends together with the opinion they support.\\
        			You must reply with the opinion you want to support.\\
        			The opinion must be reported between square brackets.}           
   		\end{itemize}
   		The adoption probability for Claude 3.5 Sonnet, GPT-4 Turbo and Llama 3 70B are shown in Fig.~\ref{fig:plot2_SI}. As it is possible to see the probability tends to be stable under change of the prompt for the most advanced models, while in less advanced models prompts seem to play a role. This is the case, for instance, of Llama 3 70B, for which 2 out of 5 prompts produce a very different adoption probability that is not well described by a tanh function. This happens also for one prompt in Claude 3.5 sonnet, but the discrepancies are less pronounced. 
    \subsection*{Role of context window length}
        \begin{figure}
				\centering
				\includegraphics[width=0.90\textwidth]{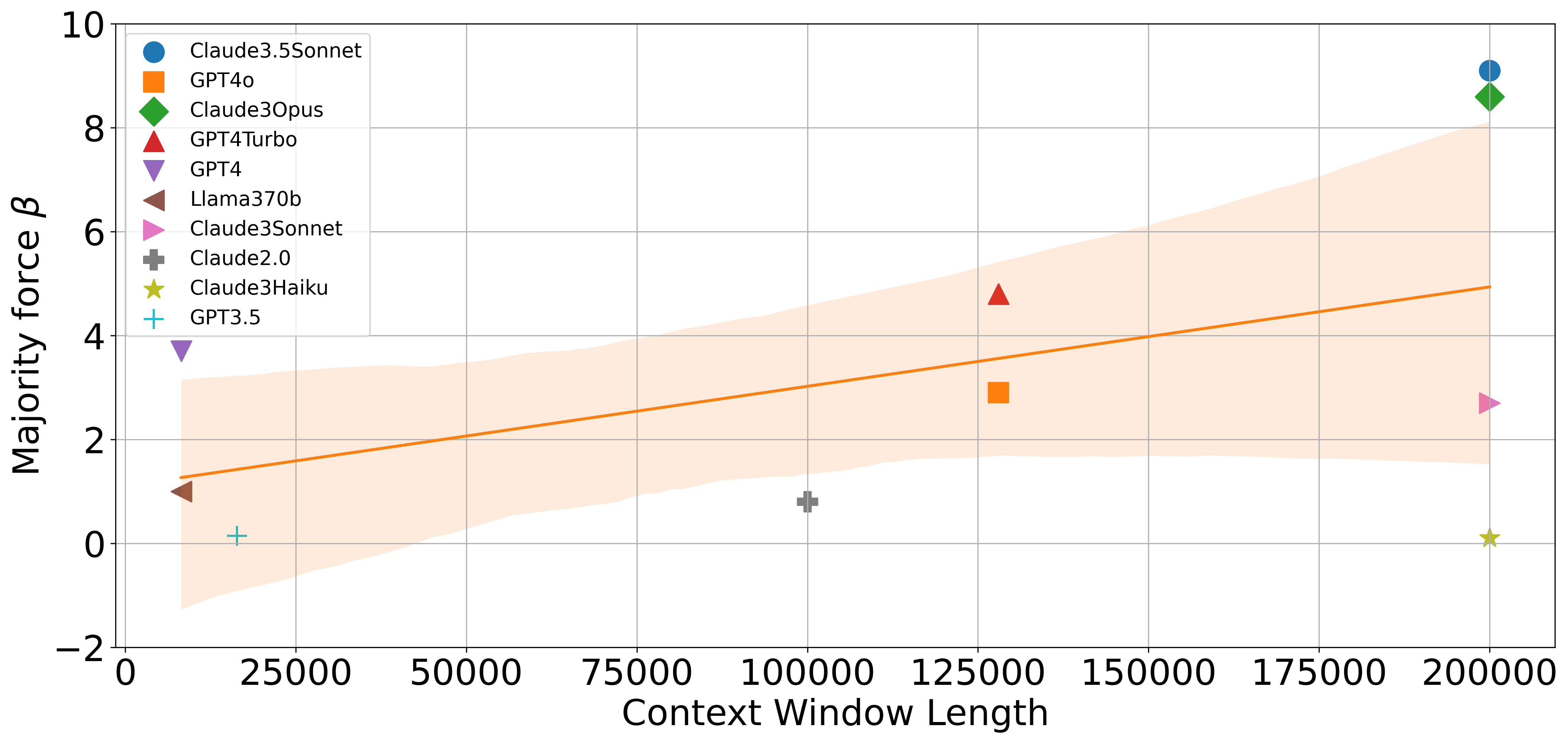}
				\caption{\textbf{Role of context window length.} Relationship between the context window length $L$, and the majority force $\beta$ for $N=50$. The two quantities show a correlation of $0.49$, a value significantly lower than the correlation between the MMLU benchmark and the majority force.}     
				\label{fig:plot3_SI}
	\end{figure}
        In order to understand if the majority force parameter $\beta$ is influenced by the language understanding and cognitive capabilities of the LLMs or rather by the context window length $L$, we repeat the analysis performed in Fig.~\ref{fig:plot2} (left panel). In this case, however, we relate the majority force with the context windows of the ten models we analyzed. As shown in Fig.~\ref{fig:plot3_SI} (left) there is a much weaker and less significant correlation ($0.49$ with a p-value of $0.15$) with respect to the MMLU benchmark, suggesting that the context window length plays a marginal role.

    \subsection*{Majority Counting}
        \begin{table}[ht]
            \centering
            \begin{minipage}{0.48\textwidth}
            \centering
            \begin{tabular}{lcc}
            \hline
            \textbf{Model} & \textbf{N=51} & \textbf{N=201} \\
            \hline
            gpt-3.5-turbo-0125 & 0.95 & 0.97 \\
            gpt-4-0613 & 0.97 & 0.97 \\
            gpt-4-turbo-2024-04-09 & 0.95 & 0.95 \\
            gpt-4o-2024-05-13 & 0.95 & 0.945 \\
            claude-2.0 & 0.94 & 0.945 \\
            claude-3-haiku-20240307 & 0.95 & 0.95 \\
            claude-3-sonnet-20240229 & 0.95 & 0.965 \\
            claude-3-opus-20240229 & 0.985 & 0.99 \\
            claude-3-5-sonnet-20240620 & 0.945 & 0.98 \\
            meta-llama-3-70b-instruct & 0.96 & 0.955 \\
            \hline
            \end{tabular}
            \caption{Success rates for majority counting (easy task)}    
            \label{tab:counting_easy}
            \end{minipage}
            \hfill
            \begin{minipage}{0.48\textwidth}
            \centering
            \begin{tabular}{lccc}
            \hline
            \textbf{Model} & \textbf{N=51} & \textbf{N=201} & \textbf{N=501} \\
            \hline
            gpt-3.5-turbo-0125 & 0.84 & 0.85 & 0.80 \\
            gpt-4-0613 & 0.89 & 0.845 & 0.85 \\
            gpt-4-turbo-2024-04-09 & 0.735 & 0.79 & 0.75 \\
            gpt-4o-2024-05-13 & 0.825 & 0.795 & 0.76 \\
            claude-2.0 & 0.75 & 0.715 & 0.715 \\
            claude-3-haiku-20240307 & 0.76 & 0.715 & 0.76 \\
            claude-3-sonnet-20240229 & 0.805 & 0.735 & 0.74 \\
            claude-3-opus-20240229 & 0.865 & 0.95 & 0.925 \\
            claude-3-5-sonnet-20240620 & 0.825 & 0.90 & 0.93 \\
            meta-llama-3-70b-instruct & 0.805 & 0.775 & 0.745 \\
            \hline
            \end{tabular}
            \caption{Success rates for majority counting (hard task)}
            \label{tab:counting_hard}
            \end{minipage}
        \end{table}

        A final relevant aspect to investigate is whether the experiments we performed could be simply related to the majority counting abilities of LLMs or if instead there is a role played by the social aspect of the simulations. In order to test this we reformulate our prompt in order to phrase it in terms of a majority counting problem. 
        \begin{itemize}
   			\item \textbf{Prompt Majority}\\
   			\textit{Reply with the most frequent letter appearing in the following sequence. The reply must be reported between square brakets.}
        \end{itemize}
        We used letters $k$ and $z$ and considered lists of different lengths up to $N=501$. We followed two different procedures for generating the lists:
        \begin{itemize}
            \item Easy task. We first generate a random fraction $p$ between $0$ and $1$ and we then generate a list containing a fraction $p$ of letters $k$ and the remaining $z$. We consider $N=51$ and $N=201$;
            \item Hard task. We first generate a random fraction $p$ between $0.4$ and $0.6$ and we then generate a list containing a fraction $p$ of letters $k$ and the remaining $z$. In this case most examples will be almost balanced making it harder for the LLM to identify the most frequent letter. For this harder task we also consider longer lists $N=51, 201, 501$.
        \end{itemize}
        The results of this analysis are reported in Tab.~\ref{tab:counting_easy} and Tab.~\ref{tab:counting_hard}. The results show substantial differences with respect to the prompt we used in our simulations. First, even models that have a very low majority force and are unable to follow the majority in systems as small as $20$, like GPT 3.5 Turbo or Claude 3 Haiku, perform very well in the task. They manage to identify the majority almost always in the easy task and they perform very well also on the harder task, even when $N=501$. For instance GPT 3.5 Turbo on the hard task with $N=500$ performs better than both GPT-4 Turbo and GPT-4o. Moreover, increasing the system size does not lead to rapid degradation of the performance, as in the case of the ``Social'' prompt. There are models that perform actually better in the longer lists, like Claude 3.5 Sonnet. These results are a strong indication that the effects we observed are derived from the social nature of the prompt and not just from the performance limitations of the LLMs. 
    \subsection*{Details on LLMs performances and benchmarks}
        \begin{figure}
				\centering
				\includegraphics[width=0.90\textwidth]{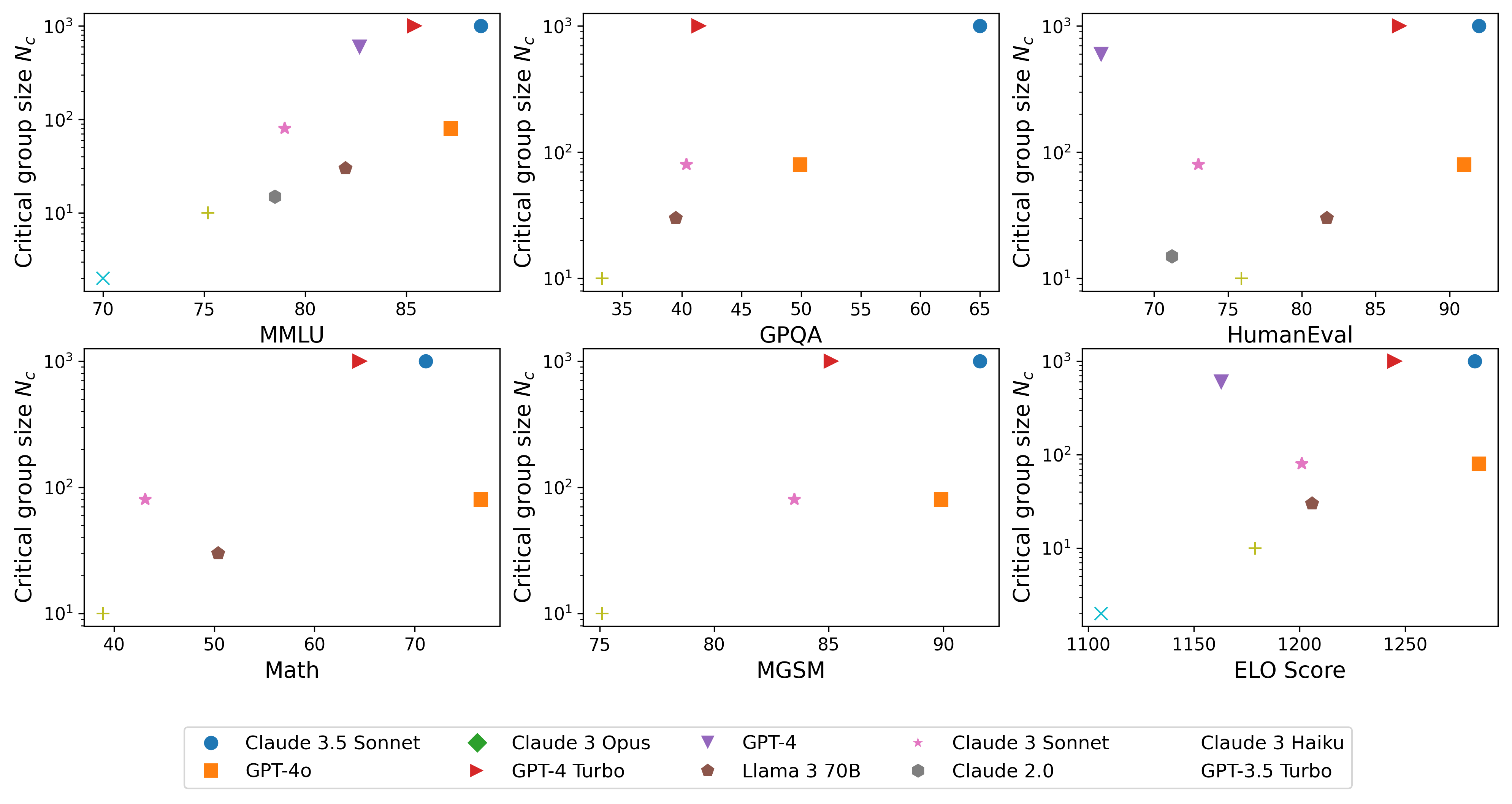}
				\caption{\textbf{Critical group size vs Benchmarks.} We compare the critical group size $N_c$ with six different benchmarks measuring the ability of LLMs. $N_c$ shows an exponential correlation with the model capabilities.}     
				\label{fig:plot4_SI}
	\end{figure}
         The performance of LLMs on the MMLU benchmark (and other tasks) has been taken from the official model information and technical reports. When this information was missing and when possible, we integrated it with independent analysis \cite{zheng2023gpt}\\
        \textbf{ChatGPT:}
        \begin{itemize}
            \item \url{https://github.com/openai/simple-evals?tab=readme-ov-file#benchmark-results}
        \end{itemize}
        \textbf{Claude:}
        \begin{itemize}
            \item \url{https://www.anthropic.com/news/claude-2}
             \item \url{https://www.anthropic.com/news/claude-3-family}
              \item \url{https://www.anthropic.com/news/claude-3-5-sonnet}
        \end{itemize}
        \textbf{Llama:}
        \begin{itemize}
            \item \url{https://huggingface.co/meta-llama/Meta-Llama-3-70B }
             \item \url{https://huggingface.co/meta-llama/Llama-3.1-70B}
        \end{itemize}

        We show in Fig.~ a comparison between six different benchmarks and the critical group size. These are: MMLU, GPQA, HumanEval, Math, MGSM, Elo Score. In all cases we observe an exponential tendency, with the critical group size growing exponentially as the performances of the model increase.
    \subsection*{Critical behavior of Curie Weiss Models}
        We can derive the critical behavior of the CW model expanding the self-consistent equation $m^*=\tanh(\beta m^*)$ for small values of the equilibrium magnetization. This allows to study the system close to the critical point. By expanding the hyperbolic tangent for small values of $ \beta m^*$
        \[
        \tanh(\beta m) = \beta m - \frac{\beta^3m^3}{3}  +  \ldots,
        \]
        which, inserted into the self-consistent equation, yields
        \[
        \ m^*\left(1-\beta +\frac{\beta^3(m^*)^2}{3}\right) = 0
        \]
        The value $m^*=0$ is always a solution of this equation, while other real solutions exist only if
        \[
            1-\beta +\frac{\beta^3(m^*)^2}{3}=0 \ \to \ \beta>\beta_c=1
        \]
        From this expansion we can also easily obtain the magnetization close to the transition point 
        \[
            1-\beta +\frac{\beta^3(m^*)^2}{3} = 0 \ \to \ m^* = \sqrt{\frac{3(\beta-1)}{\beta^3}}
        \]
        and since $\beta$ is close to $\beta_c=1$ we can further simplify the expression to 
        \[
            m^* = \sqrt{3(\beta-1)}.
        \]
    \subsection*{Derivation of $\beta_t(N)$}
        In order to write the stochastic differential equation governing the evolution of the collective opinion $m$, we define the right and left transition probabilities $R(m)$ and $L(m)$. The former gives the probability for the collective opinion to increase by an infinitesimal value $\delta m=2/N$ in a single infinitesimal update $\delta t = 1/N$; analogously the latter gives the probability of $m$ to decrease of the same quantity in the same infinitesimal interval. From the adoption probability $P(m)=0.5+0.5\tanh(\beta m)$ we can write 
        \[
            R(m)=\frac{N_-}{N}P(m)=\frac{1-m}{2}\frac{1}{2}\left[1+\tanh(\beta m)\right],
        \]
        where the first factor gives the probability of selecting an agent supporting the second opinion and the second one accounts for the probability of this AI agent to transition to the first opinion. Analogously 
        \[
            L(m)=\frac{1+m}{2}\left[1-P(m)\right]=\frac{1+m}{2}\frac{1}{2}\left[1-\tanh(\beta m)\right].
        \]
        Using $R(m)$ and $L(m)$ we can derive the drift $\nu_m$ governing the evolution of the average value of the collective opinion 
        \[
            \frac{d \langle m \rangle}{dt}=\nu_m.
        \]
        It is easy to show that the drift satisfies
        \begin{equation}
            \nu_m = \frac{\delta m}{\delta t}\left[R(m)-L(m)\right]=2P(m)-m-1=\tanh(\beta m) - m.
            \label{eq:met_drift}
        \end{equation}
        
        Analogously we can compute $D_m$, the diffusion coefficient of the collective opinion, that satisfies 
        \[
            D_m=\frac{1}{2}\frac{(\delta m)^2}{\delta t}\left[R(m)+L(m)\right].
        \]
        Using the expressions for $R(m)$ and $L(m)$, te last equation gives 
        \begin{equation}
            D_m=\frac{1}{N}[1-m\tanh(\beta m)].
            \label{eq:met_diffusion}
        \end{equation}
        We are interested in studying fluctuations around the equilibrium solution $m^*$ that satisfies the self-consistency equation
        \begin{equation}
            m^*=\tanh(\beta m^*)
            \label{eq:SI_self_consistent}
        \end{equation}
        
        Expanding \eqref{eq:met_drift} for small $\delta m=m-m^*$ we find
        \[
            \nu_m \approx -\delta m\left(1-\beta\text{sech}^2(\beta m^*)  \right) \approx -(m-m^*),
        \]
        where the last approximate equality holds already for $\beta=2$  (for which $m^* \approx 0.96$).
        
        With regard to the diffusion coefficient, the expansion of the hyperbolic tangent for large values of its argument
        \[
            \tanh(x)\approx 1-2\text{e}^{-2x}.
        \]
        inserted into \eqref{eq:met_diffusion} yields at leading order 
        \[
            D_{m}\approx\frac{2}{N}\text{e}^{-2\beta}\equiv D
        \]
        Hence the drift is approximately a linearly restoring force, while the diffusion coefficient is approximately constant. The behavior of the collective opinion around the equilibrium value can therefore be described by an Ornstein–Uhlenbeck process
        \[
            dm=-(m-m^*)dt+\sqrt{2D}dW(t),
        \]
        where $W(t)$ is a standard Wiener process.
        The variance of $m$ then is 
        \begin{equation}
            \text{Var}[m]\approx \frac{D}{\gamma}=\frac{2}{N}\text{e}^{-2\beta}.
            \label{eq:met_variance_m}
        \end{equation}
        
        The condition for $\beta_t$ is
        \begin{equation}
            m^*(\beta_t)+\sqrt{\text{Var}[m](N, \beta_t)}=1.
            \label{eq:betat}
        \end{equation}
        Expanding the self-consistent equation \ref{eq:SI_self_consistent} for $m^*\approx 1$ we obtain
        \[
            m^*\approx 1-2\text{e}^{-2\beta},
        \]
        that inserted into~\eqref{eq:betat} yields
        \[
            1-2\text{e}^{-2\beta_t}+\sqrt{\frac{2}{N}}\text{e}^{-\beta_t}=1.
        \]
        Solving for $\beta_t$ this gives
        \begin{equation}
        \beta_t\approx\frac{1}{2}\log{N}.
        \label{eq:betatvsN}
        \end{equation}

        The existence of a phase transition in the CW model can also be derived directly from the time evolution of the magnetization. 
        An ordered state, and eventually a full coordination, can emerge only if the drift is positive for small values of the magnetization, so that the disordered state $m=0$ is unstable. Expanding for small $\beta m$ we get 
        \[
            \nu_m\approx m(\beta - 1),
        \]
        from which the existence of a phase transition in $\beta_c=1$ follows immediately.
\end{document}